\documentclass[english,aps,prl,showpacs,amsmath,amssymb,twocolumn]{revtex4}
\usepackage[T1]{fontenc}
\usepackage[latin1]{inputenc}
\usepackage{graphicx}

\makeatletter

\usepackage{dcolumn}

\usepackage{bm}

\usepackage{epsf}

\usepackage{babel}
\makeatother
\begin{document}

\title{Stochastic extinction of epidemics in large populations and role
of vaccinations}

\author{Alexandra S. Landsman$^{1}$ and Ira B. Schwartz$^{1}$}

\affiliation{$^{1}$US Naval Research Laboratory, Code 6792, Nonlinear Systems
Dynamics Section, Plasma Physics Division, Washington, DC 20375.}  
%\email{<alandsma@cantor.nrl.navy.mil>}
\begin{abstract}
We investigate stochastic extinction in an epidemic model and the
impact of random vaccinations in large populations formulated in terms
of an optimal escape path. We find that different random vaccination
strategies can have widely different results in decreasing expected time
till extinction, for the same total amount of vaccines used.  
Vaccination
strategies are considered in terms of two parameters: average frequency
of vaccinations, given by $\gamma$, and the amplitude of the vaccinations,
$\epsilon$, where $\epsilon \ll 1$ refers to the proportion of the population
being vaccinated at some particular instant.  
It is found that while the average
number of individuals vaccinated per unit time, 
$\gamma \epsilon$, is kept constant, the
particular values of $\gamma$ and $\epsilon$ 
can play a highly significant role in increasing the
chance of epidemic extinction.  The findings suggest that expected time
till extinction can be significantly shortened if less frequent
vaccinations occur in larger groups, corresponding to low $\gamma$,
high $\epsilon$ strategy.
\end{abstract}
\maketitle
While deterministic  and network models have suggested certain strategies for 
epidemic control and prevention  \cite{Anderson91,moreno2002}, stochastic
models are needed to account for many of the
important features of epidemics, not the least of which is stochastic
extinction \cite{verdaska05,keeling04,doeringetal2005}. From the general theory of
finite Markov chains \cite{bartlett49}, it was shown that in stochastic
models the probability of extinction is equal to one in the asymptotic
time limit\textbf{.} Numerical \cite{west97,billings02,citeulike:cumminhgs2005}
and analytic \cite{jacquez93,allen00} comparisons of stochastic and
deterministic models have been performed and confirmed that extinction
was inevitable in the presence of stochastic effects. The numerical
results hold for very small amplitude noise as well as real finite
noise. This is in contrast to deterministic SIS or SIR models which
result in equilibrium endemic presence of infectives \cite{Anderson91}
for an appropriate choice of parameters. It is clear that stochastic
effects may result in very different dynamics from deterministic models,
particularly when extinctions occur. Since time delivery of vaccines
into populations is often
not a synchronized process\textbf{,} it is important
to take stochastic effects into account when considering different
vaccination strategies and their impact on extinction
\cite{keeling04,citeulike:schwartz04,citeulike:agur03}. Some numerical
comparisons exist on the impact of pulsed versus random vaccination
strategies in increasing the probability of stochastic extinction
\cite{keeling04}.  However, analytic comparisons of the role of various random
vaccination strategies in stochastic extinction in large populations
have not been previously carried out.

Here we use a deterministic SIS model, with small standard deviation
Gaussian noise added to model the effect of stochastic fluctuations
on the dynamics of a large population. We then consider the effect
of different random vaccination strategies on the probability of extinction
in large populations. Our approach differs from the previously used
Markov models of stochastic epidemics. These models have used asymptotic
approximations to obtain mean extinction times and quasi-stationary
distributions \cite{allen00,kryscio89,nasell96,norden82}. Our method
is based on finding the optimal escape path that occurs whenever a
system experiences a large and rare stochastic fluctuation from its
equilibrium state \cite{dykman79,dykman92,wentzell76}. In this case,
the probability densities of different trajectories during extinction
are very different, with highest probability of extinction 
occurring along the optimal
escape path. The problem of a large fluctuation for stochastic Markov
chains was treated in \cite{wentzell76}, where an optimizing
action functional is used to find the optimal escape path for a particle
in a well. Here we apply the path integral approach to find the optimal
escape trajectory that corresponds to stochastic extinction of an
SIS epidemic when the additive noise is sufficiently small. The additive
noise, while being less realistic than multiplicative noise that is
sometimes used in disease models, allows for a more tractable model,
while retaining the same qualitative results, with regards to optimal
vaccination strategies.
In our model, we obtain a closed form solution
for the stochastic fluctuation needed to push the system
along the escape trajectory, thereby computing the probability of
extinction \cite{dykman92}. The optimal escape approach simplifies the analysis
by allowing one to construct a particular deterministic trajectory, or 
optimal escape path,
of infectives in the cases that extinction occurs. We then use the
optimal escape path to estimate further the effect that vaccinations
have on the probability of extinction from a fully endemic state.

While spatial inhomogeneities may increase stochastic fluctuations
leading to faster extinctions \cite{verdaska05}, in general the size
of the population is key in determining the expected time to extinction
\cite{keeling04}. For sufficiently large populations, normalized
fluctuations approach a Gaussian distribution and scale as one over
the square root of the population size \cite{bailey75}. We therefore
focus on the SIS model where the population size, $N$, is large,
so that the stochastic fluctuations are relatively small. We model
these stochastic effects by adding a small standard deviation Gaussian
noise term, $f(t)$, to the SIS model, \begin{equation}
\frac{dS}{dt}=\mu-\beta IS+\delta I-\mu S+f(t)\label{eq:S}\end{equation}
 \begin{equation}
\frac{dI}{dt}=\beta IS-\delta I-\mu I-f(t)\label{eq:I}
\end{equation}
where $S$ and $I$ are susceptible and infective populations, respectively,
$\delta$ is a recovery rate, $\beta$ contact rate, and $\mu$ is
the birth/death rate.  
%Equation (\ref{eq:I}) uses additive, rather than
% multiplicative noise, which more accurately models the contact rates of
%the Markovian process.  However, this should not affect the qualitative
%results, which look at the optimal combination of average vaccination
%frequency $\gamma$ and amplitude $\epsilon$, while keeping $\gamma \epsilon$,
%or the average amount of individuals vaccinated per unit time constant
%(regardless of the state of the system).

All the variables in Eqns. (\ref{eq:S}) and (\ref{eq:I}) have been
scaled by the total population size, $N$, so that $S+I=1$. For $\beta>\delta+\mu$,
the equilibrium solution, $I_{eq}=\left(\beta-\left(\delta+\mu\right)\right)/\beta$
corresponds to a stable endemic state and $I_{ex}=0$ is the unstable
disease free equilibrium (DFE) corresponding to extinction. For smaller
values of $\beta$, the epidemic dies out even in the absence of any
stochastic effects, since the DFE is asymptotically stable. The term,
$f(t)$, denotes random stochastic fluctuations between the susceptibles
and the infectives, such that the total population size is conserved.
It is assumed to be uncorrelated with zero mean and standard deviation
$D$.

The Langevin formulation, corresponding to 
Eqns. (\ref{eq:S}) and (\ref{eq:I}), most closely agrees with numerical
simulation when the disease parameters are near
threshold, where the equilibrium number of infectives, $I_{eq}$, 
is significantly smaller than susceptibles.  Numerical results, however,
show good
agreement with the Langevin or the Fokker-Planck approaches
when the equilibrium level of infectives is as
large as one third of the total population (corresponding to $\Lambda \equiv 
1/(1-I_{eq}) = 1.5$) \cite{doeringetal2005}.
The analysis in the present paper does not impose any restriction on the
parameters that determine the equilibrium levels of infectives.  There
is however a constraint on the size of stochastic fluctuations, which
should be sufficiently small for the optimal escape analysis to be
applicable. Because the size of stochastic fluctuations is determined by
the size of the total population, the present results should also 
apply for cases
that are close to the threshold, provided the total population size is
sufficiently large.  

Since the stochastic fluctuations are small, the solution will spend
most of its time close to the equilibrium state, $I_{eq}$. However,
given a sufficient amount of time, stochastic fluctuations will build
up in such a way that the number of infectives will go to $I_{ex}=0$,
leading to disease extinction. In large populations, this is a highly
improbable event, where the length of time to extinction has an exponential
dependence on the size of the population \cite{allen00,kryscio89}.
We are interested in how a string of vaccinations increases the probability
of extinction of an epidemic due to stochastic effects. In this case,
{}``vaccinations'' can also stand for various preventative strategies,
such as wearing face masks, or avoiding all contact with the infectives.
Suppose at any interval of time, $\triangle t$, there is a probability,
$\gamma\triangle t$, of an occurrence of a vaccination, whereby some
proportion of the population, $\epsilon\ll1$ is vaccinated.
The string of vaccinations is then given by a sequence of Poisson
distributed pulses of amplitude $\epsilon$ and average number of
vaccinations per unit time is $\gamma\epsilon$.

Suppose that at some time, $t=t_{j}$, a number, $N_{j}$, of individuals
are vaccinated, so that $N_{j}/N=\epsilon$ at $t=t_{j}$, and $N$
is the total number of people in the population.  If each individual
acts independently of everybody else, than the vaccination process is
Poisson distributed with $\epsilon = 1/N$.  However, it often happens that
individuals are vaccinated in groups.  One example would be
when colleges offer vaccinations to 
all of the students over some short period of time, usually involving a few
days.
In this case, different vaccination cites can act as independent agents making
a decision on when to offer group vaccinations, so that the vaccination
process has a Poisson distribution with $\epsilon=N_{j}/N$, where $N_j$
is the size of the group that a particular cite chooses to vaccinate at
time $t_j$.  For simplicity we assume that the size of the group being
vaccinated is the same at all times (so that $\epsilon$ is a constant),
however the results can be generalized to variable group sizes.

Since vaccinations confer immunity, 
the number of
these immune individuals will only change due to death, $\mu$. The
time-evolution of immune individuals is then given by, 
\begin{equation}
\zeta(t)=\sum_{j=1}^{n}N_{j}(t)/N=\sum_{j=1}^{n}\epsilon
e^{-\mu\left(t-t_{j}\right)},\label{eq:zeta}
\end{equation}
 where $\zeta(t)$ is Poisson distributed, corresponding to a string
of random vaccinations occurring at various times, $\{ t_{1},t_{2},...t_{n}\}$,
with $t_{n}<t$. Since the total population stays constant, $I(t)+S(t)+\zeta(t)=1,$
we substitute for $S$ in Eqn. (\ref{eq:I}), to get \begin{equation}
\frac{dI}{dt}=aI-\beta I^{2}+f(t)-\beta I\zeta(t),\label{eq:If}\end{equation}
 where $a=\beta-\delta-\mu$. Neglecting the last term, the above
equation corresponds to a noisy logistic model, used in a variety
of contexts, such as chemical reactions, transmission of rumors, and
population growth \cite{norden82}. Equation (\ref{eq:If})
also corresponds to the Langevin equation of a particle trapped in
an over-damped potential well centered at $I_{eq}$ with a potential
maxima at $I_{ex}$.

In the absence of vaccinations, $\zeta=0$, the Gaussian noise term,
$f(t)$ causes fluctuations about the equilibrium state. Given enough
time, random noise fluctuations will combine in such a way as to drive
the particle towards $I_{ex}$, resulting in extinction.
For small standard deviation, $D$, of the noise term, $f(t)$, this
extinction will occur along an optimal escape path \cite{dykman92,feynman65}.
For uncorrellated Gaussian noise: $<f(t)f(t\prime)>=\delta(t-t\prime)$,
the probability of optimal escape is then given by, \begin{equation}
P\lbrack I_{esc}\rbrack^{(0)}=exp\lbrack-\frac{1}{2D}\int_{-\infty}^{\infty}f_{opt}(t)^{2}dt\rbrack\equiv exp\lbrack-\mathcal{R}^{(0)}/D\rbrack,\label{eq:gauss}\end{equation}
 where $f_{opt}$ denotes the stochastic fluctuations that occur when
the trajectory moves along the optimal escape path \cite{dykman79},
for which $\mathcal{R}^{(0)}$ is minimized. It can be seen that for
sufficiently small standard deviation of noise, $f_{opt}\gg D$, the
probability of extinction along any other non-optimal trajectory becomes
negligible. The superscript on the $P\lbrack I_{esc}\rbrack^{(0)}$
and $\mathcal{R}^{(0)}$ terms in Eqn. (\ref{eq:gauss}) indicates
that that this expression is a solution for the probability of an
optimal escape path in the absence of any vaccinations. 

Since any vaccinations should increase the probability of epidemic
extinctions, using perturbation theory we will obtain an additional
correction term, $\mathcal{R}^{(1)}$, that is a direct result of
a series of random vaccinations. The optimal noise, $f_{opt}(t)$,
and the resultant optimal escape path is found by minimizing the Gaussian
noise over the escape trajectory, 
\begin{equation}
\mathcal{R}=\frac{1}{2}\int_{-\infty}^{\infty}f(t)^{2}dt=\frac{1}{2}\int_{-\infty}^{\infty}L(I,\dot{I})dt.\label{eq:prob}
\end{equation}
 This is equivalent to minimizing action over a trajectory in an Euler-Lagrange
system, where the Lagrangian is given by $L(I,\dot{I}) = \left(\dot{I}-aI+\beta
I^{2}\right)^{2}$. The optimal
escape path then corresponds to the deterministic trajectory given
by the Euler-Lagrange equations, $\partial L/\partial I=d\left(\partial
L/\partial\dot{I}\right)/dt$.  The optimal escape path is a heteroclinic
orbit connecting $I_{eq}$ to $I_{ex}$. The steady states
,$I_{eq}$ and $I_{ex}$, are both saddles in the conservative
Euler-Lagrangian system given by Eqn. (\ref{eq:prob}).  The integral in Eqns.
(\ref{eq:gauss}) and (\ref{eq:prob}) is taken for $t$ from $-\infty$
to $\infty$, which corresponds to the motion along the heteroclinic
orbit of the system connecting the saddle points. Using Euler-Lagrange
equations of motion, and conservation of energy, we solve for the
optimal escape path, $\{ I_{esc}(t),\dot{I}_{esc}(t)\},$ in the absence
of vaccinations. After solving for $\dot{I}_{esc}$ and integrating,
we get the trajectory of infectives as a function of time for the
most probable path of extinction, \begin{equation}
I_{esc}(t)=\frac{a}{\beta}\left[\frac{exp\left(a^{2}t\right)}{1+exp\left(a^{2}t\right)}\right].\label{eq:qesc}\end{equation}
Using Eqn. (\ref{eq:qesc}) and $f_{opt}(t)=\dot{I}_{esc}(t)-aI_{esc}+\beta I_{esc}^{2}$, we solve for the optimal noise:
\begin{equation}
f_{opt}(t)=\frac{2a^{2}}{\beta}\left[\frac{exp\left(a^{2}t\right)}{\left(1+exp\left(a^{2}t\right)\right)^{2}}\right].\label{eq:qesc2}\end{equation}
 Th optimal path is perturbed in the presence of vaccinations, $\zeta(t)$.
For small $\zeta$ ($\zeta\ll f_{opt}$), however, the perturbation
is small and the effect of vaccinations on the probability of extinction
can be obtained as a first order correction to $\mathcal{R}^{(0)}$
in Eqn. (\ref{eq:gauss}) \cite{dykman92}, 
\begin{equation}
\mathcal{R}^{(1)}[\zeta]=-\int_{-\infty}^{\infty}\beta I_{esc}(t)f_{opt}(t)\zeta(t)dt.\label{eq:corrp}\end{equation}
 The above equation gives a correction to $\mathcal{R}$ that can
be integrated over the Poisson probability distribution, $\mathcal{P}_{\zeta}[\zeta(t)]$,
of vaccinations to find the increase in chance extinctions. The correction
term, $\mathcal{R}^{(1)}$ in the above equation is small compared
to $\mathcal{R}^{(0)}$. However, it can still be large compared to
the standard deviation of noise, $D$ (see Eqn. (\ref{eq:gauss})),
so that vaccinations can significantly increase chance extinctions.
Intuitively, vaccinations increase the probability of extinction by
decreasing the amount of stochastic fluctuations needed to push the
infectives along the optimal escape path. 

\begin{figure}[ht]
\includegraphics[width=2.75in,keepaspectratio]{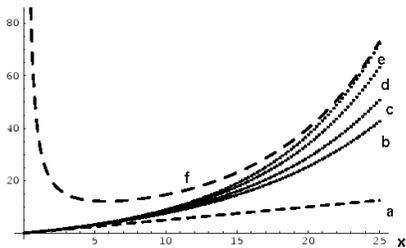} 
\caption{Scaled Logarithmic increase in extinction probability, 
$\Xi$, as a function of $x$.  (a) Approximation for
low $x$, given by Eqn. (\ref{eq:suscepsmall}), (b) $\tilde{\mu} = 0.7$,
(c) $\tilde{\mu} = 1$, (d) $\tilde{\mu} = 2$, (e) Limit of large $\tilde{\mu}$,
given by Eqn. (\ref{eq:suscep}), (f) Asymptotics valid for large $\tilde{\mu}$
  and large $x$, given by Eqn. (\ref{eq:expansion}).}
\label{fig:Escape} 
\end{figure}

To find the increase in probability of extinction due to a probabilistic
sequence of vaccination pulses, we evaluate $\mathcal{R}^{1}[\zeta]$
with respect to different realizations of $\zeta(t)$. The probability
of epidemic extinction along the optimal path in the presence of vaccinations
is then given by \cite{dykman92}, \begin{equation}
P[I_{esc}]=P_{0}\int_{-\infty}^{\infty}\exp\left(-\mathcal{R}^{(1)}/D\right)\mathcal{P}_{\zeta}[\zeta(t)]\mathcal{D}\zeta(t),\label{eq:newprob}\end{equation}
 where $P_{0}=P\lbrack I_{esc}\rbrack^{(0)}$.
We are interested in a sequence of vaccinations given by a Poisson
distribution, with a specific realization given by Eqn. (\ref{eq:zeta}).
For a Poisson distribution, the probability density, $\mathcal{P}_{\zeta}[\zeta(t)]\mathcal{D}\zeta(t)$,
of any specific realization of $n$ random vaccinations over the extinction
interval, $2T$, is given by: $dt_{1}/2T...dt_{n}/2T$. Using Eqns.
(\ref{eq:zeta}) and (\ref{eq:qesc})- (\ref{eq:newprob}), and rescaling
time as $t\rightarrow a^{2}t$, we get the increase in the chance
of extinction, conditional on $n$ vaccinations of amplitude $\epsilon$,
\begin{equation}
\frac{P[I_{esc}|n]}{P_{0}}=\int_{-\tilde{T}}^{\tilde{T}}exp\left(\tilde{\mu}x\int\phi(t)\sum_{j=1}^{n}
 e^{-\tilde{\mu}\left(t-t_{j}\right)}dt\right)\frac{dt_{1}}{2\tilde{T}}...\frac{dt_{n}}{2\tilde{T}}\label{eq:prob_full}\end{equation}
 where $\tilde{\mu}=\mu/a^{2}$, $\tilde{T}=a^{2}T$, $x$ is a function
of parameters, \begin{equation}
x\equiv\frac{2a^{3}}{\beta^{2}}\left(\frac{\epsilon\beta}{\mu D}\right)=\frac{2\epsilon}{D}\left(\frac{\left(\beta-\delta-\mu\right)^{3}}{\mu\beta}\right)\label{eq:x}\end{equation}
 and $\phi(t)$ is the scaled $I_{esc}(t)f_{opt}(t)$ variable obtained
from Eqns. (\ref{eq:qesc}) and (\ref{eq:qesc2}), \begin{equation}
\phi\left(t\right)=\frac{exp\left(2t\right)}{\left(1+exp\left(t\right)\right)^{3}}.\label{eq:phi}\end{equation}
 Since the occurrence of any of the pulses over the interval is independent
of the other pulses, Eqn. (\ref{eq:prob_full}) can be rewritten 
as \cite{feynman65},
\begin{equation}
\frac{P[I_{esc}|n]}{P_{0}}=\left(\int_{-\tilde{T}}^{\tilde{T}}
exp\left(\tilde{\mu}x\int\phi(t)e^{-\tilde{\mu}\left(t-s\right)}dt\right)\frac{ds}{2\tilde{T}}\right)^{n}.\label{eq:prob_full2}\end{equation}
 The above equation gives the probability of escape when the number
of pulses during the escape interval is $n$. Using a Poisson probability
distribution, and summing over all possible $n$, the total escape
probability is then given by, \begin{equation}
\frac{P[I_{esc}]}{P_{0}}=\sum_{n}A^{n}\frac{\bar{n}^{n}}{n!}e^{-\bar{n}},\label{eq:poisson}\end{equation}
 where $A^{n}$ is given by the right hand side of Eqn. (\ref{eq:prob_full2}).
If the average number of pulses per unit time is $\gamma$, then $\bar{n}=2\gamma T$
in the above equation. The sum in Eqn. (\ref{eq:poisson}) is an expansion
of the exponential function, $\exp\left[-\left(1-A\right)\bar{n}\right]$.
Taking the log on both sides and scaling by
$a^2/\delta$, the LHS
is: $\Xi\equiv\frac{a^{2}}{\gamma}\ln\left(\frac{P[I_{esc}]}{P_{0}}\right)$;
and the RHS side becomes:
\begin{equation}
\Xi=
-\int_{-\infty}^{\infty}\left(1-\exp\left[\tilde{\mu}x
\int_{s}^{\infty}\phi(t)e^{-\tilde{\mu}\left(t-s\right)}dt\right]\right)ds\label{eq:poisson2}\end{equation}
In the above equation, the limits of integration, $\pm \tilde{T}$
have been extended
to infinity, since the optimal escape trajectory is along the heteroclinic
orbit.  Figure \ref{fig:Escape} plots the scaled logarithmic increase in
escape probability, $\Xi$ as a function of $x$ for $\tilde{\mu} = 0.7,
1$, and $2$, given by curves (b)-(d).  As  $\tilde{\mu}$ increases, $\Xi$
asymptotes fast to the upper limit 
given by curve (e) in the same figure \cite{note4}.
Letting $\tilde{\mu} \rightarrow \infty$ in Eqn. (\ref{eq:poisson2}),
the expression corresponding to curve (e) in Fig.  \ref{fig:Escape} is,
\begin{equation}
\Xi=-\int_{-\infty}^{\infty}\left(1-\exp\left[x\phi(s)\right]\right)ds
\label{eq:suscep}\end{equation}
$\phi(s)$ is a bounded function with a global maxima at $s_{0}=\ln2$,
and asymptotically approaching zero as $s\rightarrow\pm\infty$.
For small $x$, Eqn. (\ref{eq:suscep}) can be approximated as \begin{equation}
\Xi\sim\int_{-\infty}^{\infty}\phi(s)ds\approx\frac{x}{2};\qquad
x\rightarrow0\label{eq:suscepsmall}\end{equation}
This line, $\Xi = x/2$, is plotted in Fig.  \ref{fig:Escape}, curve (a) 
and provides a good
approximation for all values of $\tilde{\mu}$, while $x$ is sufficiently
small.  Since $x$ has a linear dependence on the vaccination amplitude,
$\epsilon$, and
$\Xi\equiv\frac{a^{2}}{\gamma}\ln\left(\frac{P[I_{esc}]}{P_{0}}\right)$ is scaled by
the average vaccination frequency, $\gamma$, we have, using Eqns.
(\ref{eq:x}), and (\ref{eq:suscepsmall}), \begin{equation}
\ln\left(\frac{P[I_{esc}(t)]}{P_{0}}\right)\propto\epsilon\gamma;\qquad
x\rightarrow 0\label{eq:prop}\end{equation}
From Eqn. (\ref{eq:prop}), it is clear that increase in extinction probability
has an exponential dependence on the average number  of
vaccinations per unit time, given by $\epsilon \gamma$.  However, this is only
true in the range of smaller $x$, whereby the linear approximation plotted
in  Fig.  \ref{fig:Escape}, curve (a) is valid.  At higher $x$, $\Xi$ has
a nonlinear dependence on $x$ (see Fig.  \ref{fig:Escape}), 
suggesting that increasing $\epsilon$ will be more effective in increasing
the extinction probability than increasing $\gamma$.  In this range,
it is therefore not just the average vaccinations per unit time,
$\epsilon \gamma$, but the amplitude of these vaccinations, or the degree
of fluctuation about the average, that is important in decreasing time
till extinction. 

For large $x$, we can use Laplace's method \cite{bender99} to derive
an asymptotic approximation for Eqn. (\ref{eq:suscep}),
%\begin{equation}
\begin{multline}
\Xi\sim\sqrt{\frac{2\pi}{-x\phi''(t_0)}}e^{x\phi(t_0)}\times\\
\left[1+\frac{1}{x}\left(\frac{\left(d^{4}\phi/dt^{4}\right)(t_0)}{8\left[\phi''(t_0)\right]^{2}}-\frac{5\left[\phi'''(t_0)\right]^{2}}{24\left[\phi''(t_0)\right]^{3}}\right)\right],\quad
x\rightarrow\infty\label{eq:expansion}
\end{multline}
%\end{equation}
where $\phi(t_0)$ and its various derivatives are evaluated using 
Eqn. (\ref{eq:phi}) with $t_0 = \ln 2$.  

The curve given by Eqn. (\ref{eq:expansion}) is plotted in Fig.
\ref{fig:Escape}, curve (f).  As can be seen from the figure, 
Eqn. (\ref{eq:expansion})
gives an upper limit on the increase in extinction probability due to
vaccinations.  From Eqn. (\ref{eq:expansion}), at higher
$x$, $\ln\left(P[I_{esc}(t)]/P_{0}\right)$ has an exponential dependence
on $\epsilon$ and only a linear dependence on $\gamma$. It therefore
follows that in this range of parameters, keeping the average number
of vaccinations per unit time fixed, the strategy of delivering high
amplitude lower frequency vaccinations is much more effective in increasing
the probability of stochastic extinctions in a fully blown epidemic.
This perhaps makes sense in the context of stochastic extinctions,
when one considers that in the absence of any vaccinations, the time
till extinction depends on the size of stochastic fluctuations relative
to the size of the population. Random vaccinations can themselves
be considered as a source of positive stochastic fluctuations, adding
to the stochastic fluctuations in disease transmission. 
Since less
frequent, higher amplitude vaccinations correspond to a greater standard
deviation of vaccinations, 
they add more to the stochastic fluctuations of the whole population.
This results in lesser expected time till extinction, compared to other
random vaccination strategies that use the same number of vaccines.

The analysis in this paper suggests that depending on parameters (given
by birth, recovery and contact rates),
it can be far more effective to vaccinate
individuals in groups, rather than allowing each individual to make an
isolated decision.  This can be understood as follows:
if each person vaccinates independently of everybody else than $\epsilon$
takes the smallest possible value of $1/N$, corresponding to the amplitude of
an individual vaccination.  However, if a group of size $N_j$ vaccinates
at approximately the same time, then the amplitude, $\epsilon$ is given 
by $N_j/N$.  It follows that
even if $\gamma \epsilon$ or
the average number of individuals vaccinating per unit time
is the same in both
cases, the exptected time till extinction may be
significantly shortened.

We gratefully acknowledge support from ONR, AFMIC and ARO. 
ASL is currently a National Research Council post doctoral fellow.
%\bibliographystyle{plain} 
%\bibliographystyle{/home1/nlschaosa/schwartz/texinputs/apsrev}
%\bibliography{epidemics}

\begin{thebibliography}{10}
\bibitem{note4}
\bibinfo{note}{$\tilde{\mu}$ may be large when the epidemic rate of growth is
  comparable or smaller than the immigration or birth term.}
\end{thebibliography}

\newpage{}
\end{document}